\documentclass[showpacs,amsmath,amssymb,aps,10pt,reprint,superscriptaddress,prb]{revtex4-1}
\usepackage{graphicx}
\usepackage{bm}
\usepackage[breaklinks=true,colorlinks=true,linkcolor=blue,urlcolor=blue,citecolor=blue]{hyperref}
\usepackage{graphics}
\usepackage{dcolumn}
\usepackage{amsmath,amssymb}
\usepackage{mathdots}
\usepackage{natbib}
\usepackage{soul,color}
\usepackage{epstopdf}
\usepackage[note-name]{notes2bib}

\begin{document}

\title{Local flux-flow instability in superconducting films near $T_\mathrm{c}$}
    \author{Alexei I. Bezuglyj}
\affiliation{Physics Department, V. Karazin Kharkiv National University, 61077 Kharkiv, Ukraine}
\affiliation{Institute for Theoretical Physics, NSC-KIPT, 61108 Kharkiv, Ukraine}
    \author{Valerij~A.~Shklovskij}
\affiliation{Physics Department, V. Karazin Kharkiv National University, 61077 Kharkiv, Ukraine}
    \author{Ruslan V. Vovk}
    \author{Volodymyr~M. Bevz}
\affiliation{Physics Department, V. Karazin Kharkiv National University, 61077 Kharkiv, Ukraine}
    \author{Michael Huth}
\affiliation{Physikalisches Institut, Goethe University, 60438 Frankfurt am Main, Germany}
    \author{Oleksandr V. Dobrovolskiy}
\email[Corresponding author: ]{Dobrovolskiy@Physik.uni-frankfurt.de}
\affiliation{Physikalisches Institut, Goethe University, 60438 Frankfurt am Main, Germany}
\affiliation{Physics Department, V. Karazin Kharkiv National University, 61077 Kharkiv, Ukraine}

\begin{abstract}
Larkin and Ovchinnikov established that the viscous flow of magnetic flux quanta in current-biased superconductor films placed in a perpendicular magnetic field can lose stability due to a decrease in the vortex viscosity coefficient $\eta$ with increasing velocity of the vortices $v$. The dependence of $\eta $ on $v$ leads to a \emph{nonlinear} section in the current-voltage ($I$-$V$) curve which ends at the flux-flow instability point with a voltage jump to a highly resistive state. At the same time, in contradistinction with the nonlinear conductivity regime, instability jumps often occur in \emph{linear} $I$-$V$ sections. Here, for the elucidation of such jumps we develop a theory of \emph{local} instability of the magnetic flux flow occurring not in the entire film but in a narrow strip across the film width in which vortices move much faster than outside it. The predictions of the developed theory are in agreement with experiments on Nb films for which the heat removal coefficients and the inelastic scattering times of quasiparticles are deduced. The presented model of local instability is also relevant for the characterization of superconducting thin films whose performance is examined for fast single-photon detection.
\end{abstract}

\maketitle

\section{Introduction}\label{s1}

The transition of a current-carrying thin-film superconductor to the resistive state is widely used as an efficient means for the detection of electromagnetic radiation. This resistive response of superconductors is exploited in bolometric transition-edge sensors \cite{Ric94jap,Gil99apl,Shi17tas,Niw17nsr}, and currently great efforts are directed at a further improvement of the performance of single-photon detectors
\cite{Gol01apl,Sem01pcs,Sem05pjb,Had09nph,Zot12prb,Bul12prb,Nat12sst,Eng15sst,Vod17pra,Kor18pra}. In this regard, thin-film superconductor single-photon detectors have a series of advantages over detectors based on, e.g. tunnel junctions \cite{Twe86epl,Bar88boo,Nak93jap,Par05sst}. Among these are the technological simplicity of both, the detecting and the read-out devices, a broad spectral sensitivity, high photon count rates ($\gtrsim 1$\,GHz), and high efficiencies of quantum detection \cite{Nat12sst,Eng15sst}. Accordingly, the elucidation of quasiparticle energy relaxation mechanisms in the nonequilibrium state induced by large dc currents
\cite{Leo11prb,Emb17nac,Cap17apl,Kog18prb}, high ac frequencies \cite{Pom08prb,Wor12prb,Che14apl,Lar17pra,Tik18prb,Dob19rrl,Loe19acs} or appearing in consequence of photon absorption \cite{Gol01apl,Sem01pcs,Sem05pjb,Had09nph,Zot12prb,Bul12prb,Nat12sst,Eng15sst,Vod17pra,Kor18pra} has become a matter of intensive research both, experimentally
\cite{Pom08prb,Leo11prb,Wor12prb,Che14apl,Lar17pra,Emb17nac,Cap17apl,Dob17sst,Kor18pra,Loe19acs} and theoretically \cite{Sem05pjb,Pom08prb,Zot12prb,Bul12prb,Eng15sst,Vod17pra,Kog18prb,Tik18prb,Qia18nsr}.

The electric response of a thin-film detector to an absorbed quantum of electromagnetic radiation is associated with local heating of the superconducting film. Absorption of a photon with energy $\hbar \omega$ leads to the creation of an electron with the energy $E\sim\hbar \omega \gg \Delta$, where $\Delta$ is the  gap  in the energy spectrum of the quasiparticles. The relaxation of the high-energy electrons leads to the appearance of a cloud of non-equilibrium (hot) quasiparticles. The number of such quasiparticles is of the order of $\hbar \omega /\Delta \gg 1$, and they are formed at a hot spot where superconductivity is locally suppressed. This causes a redistribution of the bias current and, consequently, an increase of the current density in the adjacent superconducting areas. In absence of a magnetic field, an increase of the current density can result in two distinct scenarios of the electrical resistance to appear \cite{Sem05pjb,Zot12prb,Bul12prb,Eng15sst,Vod17pra}. In wider films, vortex-antivortex pairs are formed in the hot spot region and these pairs are driven towards the opposite edges of the bridge under the action of the Lorentz force. In narrower films (or at higher photon energies), superconductivity can be completely destroyed across the entire width of the bridge, that can be imagined as a normal conducting domain crossing the film. The formation of both, vortex-antivortex pairs and normal domains in the bridge leads to the appearance of a voltage drop that can be registered by a read-out device. Since both these resistivity mechanisms do not include diffusion of non-equilibrium quasiparticles over appreciable large distances, the photon count rate in thin-film detectors is usually by two to three orders of magnitude higher than in tunnel junction detectors where such a diffusion is involved.

In a perpendicular magnetic field, the transition of a current-carrying superconducting film to the normal state is often mediated by the flux-flow instability
\cite{Lar75etp,Lar86inb,Bez92pcs,Mus80etp,Kle85ltp,Vol92fnt,Xia99prb,Per05prb,Att12pcm,Sil12njp,Gri15prb,Shk17prb}, the microscopic theory of which was developed by Larkin and Ovchinnikov \cite{Lar75etp, Lar86inb}. This non-equilibrium state is produced by the electric field at the core of vortices instead of being photon-induced by the formation of a current assisted hot spot, and it is this resistivity mechanism which will be dealt with in this paper. The flux-flow instability causes dc-assisted quenching \cite{Wor12prb,Che14apl,Dob15apl} of microwave transmission lines and contains information on the quasiparticle relaxation \cite{Per05prb,Leo11prb,Att12pcm,Cap17apl}. In return, the relaxation of quasiparticles is pivotal in almost all phenomena harbouring non-equilibrium superconductivity \cite{Gra81boo,Leo11prb,Cir11prb,Wor12prb,Bec13prl,Vis14prl,Lar15nsr,Dob19rrl}, and it is in particular crucial for photon detection \cite{Had09nph,Kor15tas,Cap17apl,Vod17pra} and optical control of dynamical states \cite{Mad18sca}. For this reason, the understanding of mechanisms of the flux-flow instability is highly relevant for the optimization of the superconducting devices' performance.

The physical cause of the flux-flow instability is associated with a decrease in the number of quasiparticles in the vortex cores under the action of an electric field. In particular, the decrease in the number of quasiparticles leads to a shrinkage of the vortex cores and a decrease in the vortex viscosity coefficient with increase of the vortex velocity. In consequence of this, the viscous force has a maximum as a function of the vortex velocity, and as soon as the Lorentz force exceeds this maximum the viscous flow of the vortices becomes unstable. As it was shown later in Ref.~\cite{Bez92pcs}, the occurrence of the flux-flow instability cannot be considered separately from heating of the superconductor caused by the viscous flux flow. Only when this overheating is taken into account a quantitative agreement between the theory of flux-flow instability and experiments can be obtained \cite{Bez92pcs,Xia98prb,Xia99prb,Per05prb,Cir11prb}.

In this work, we consider a \emph{local} instability of the vortex motion in a stripe (channel) of width $\delta$ crossing the superconducting bridge. In this case, care should be applied when deducing physical quantities from $I$-$V$ curves, as the flux-flow instability takes place not in the entire superconducting bridge, but only in the narrow stripe. In the course of our analysis, we will find the $\delta$-dependent current density $J^* (\delta)$ at which the flux flow becomes unstable leading to a strong \emph{local} heating of the superconductor. The importance of $J^*$ for photon detection becomes apparent as follows: If $J^* > J_\mathrm{eq}$, where $J_\mathrm{eq}$ is the current density corresponding to the equilibrium of the nonisothermal $N/S$-boundary (see Refs. \cite{Gur84spu,Bez84ltp} and references therein), then the entire film can transit into the normal state due to the growth of the normal conducting domain. By contrast, if $J^* < J_\mathrm{eq}$, a non-stationary normal domain can appear in the film, which first grows, but after attaining its maximal size begins to shrink and eventually vanishes \cite{Bez84ltp}. Thus, for the reliable recovery of the superconducting state of the film after pulsed local heating, the transport current density $J$ should be smaller than $J_\mathrm{eq}$.

In what follows we analyze one of the initial stages of the formation of the resistive state in a thin-film superconducting bridge in consequence of the flux-flow instability. We adopt the theory developed in Refs. \cite{Lar75etp,Lar86inb,Bez92pcs} for the local instability occurring in a narrow stripe across the superconducting film and argue that in this case the $I$-$V$ curves maintain a linear shape up to the instability point at which renormalized heat removal parameters can be deduced. Our theoretical predictions are in agreement with experimental data on Nb films for which the heat removal coefficients and the inelastic scattering
times of quasiparticles are deduced. The presented model of local instability is relevant for the characterization of superconducting thin films whose performance is examined for fast single-photon detection.

The paper is organized as follows. In Sec. \ref{s2} we present the theoretical model and calculate the temperature distribution in the film containing a stripe of width $\delta$ with moving vortices. In Sec. \ref{s3}, a local flux-flow instability is considered with account for a finite rate of heat removal. Section \ref{s4} is devoted to experimentally measured $I$-$V$ curves on Nb films with different morphology. In Sec. \ref{s5}, we discuss the experimental results in comparison with the developed theory. Conclusions round up our presentation in Sec. \ref{s6}.

\section{Theoretical model}
\subsection{Temperature distribution in a film containing a stripe of mobile vortices}
\label{s2}

\begin{figure}[t!]
    \includegraphics[width=1\linewidth]{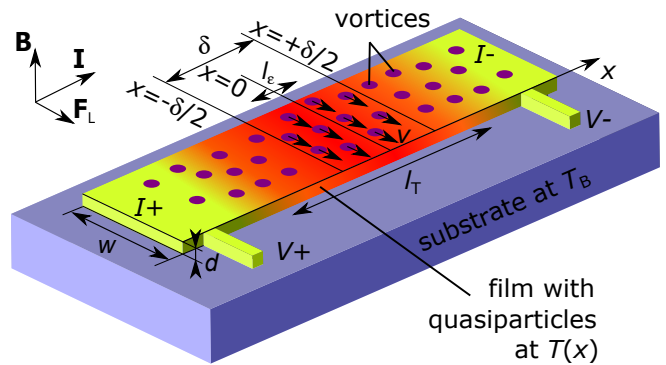}
    \caption{Geometry of the problem. A superconducting film of thickness $d$ and  width $w$ is in a perpendicular magnetic field with induction $\mathbf{B}$. Due to defects, the arrangement of vortices in the film deviates from the perfect hexagonal lattice. In the theoretical model, only vortices in a narrow stripe of width $\delta$ can move across the film with velocity $\mathbf{v}$ under the action of the Lorentz force $\mathbf{F}_\mathrm{L}$ induced by the transport current $\mathbf{I}\parallel\mathbf{x}$. The vortices outside of the stripe are assumed to be immobile. Due to local overheating caused by escaping quasiparticles from the vortex cores, the quasiparticle temperature $T$ in and near the stripe becomes larger than the substrate temperature $T_\mathrm{B}$, as indicated by the color gradient.}
    \label{fig1}
\end{figure}

We consider a superconducting film of width $w$ and thickness $d$ in a perpendicular magnetic field  with induction $\mathbf{B}$, see Fig. \ref{fig1}. In the film, we assume a stripe of width $\delta$ in which vortices can move across the film with velocity $\mathbf{v}$ under the action of the Lorentz force $\mathbf{F}_\mathrm{L}$ induced by a transport current $\mathbf{I}\parallel\mathbf{x}$. The vortices outside of the stripe are assumed to be immobile. This assumption, as well as the general justification of the model, will be discussed further in Sec. \ref{s5}. The theoretical task is to calculate the quasi one-dimensional quasiparticle temperature distribution $T(x)$ in the film containing the stripe of mobile vortices and then, to determine the voltage and the current at the instability point with account for finite heat removal from the film into the substrate.

The voltage drop $V$ along the film is determined by the average of the time derivative of the phase difference  of the superconducting order parameter at the film edges
\begin{equation}\label{1}
    V = \frac{\hbar}{2e_0}\overline{\frac{d\varphi}{dt}},
\end{equation}
where $e_0$ is the electron charge. Each time the film is crossed by a vortex the phase difference changes by $2\pi$. The vortex density in the film is given by $n=B/\phi_0$, which is why in consequence of the vortex motion with velocity $v$ the phase difference in the stripe of width $\delta$ is changed by $2\pi B v\delta/\phi_0$ per unit of time. With the definition of the magnetic flux quantum $\phi_0=\pi \hbar c/e_0$ the voltage on the film can be written as $V=B v\delta/c$, where $c$ is the speed of light. The electric field in the stripe follows from the relation $E = V/\delta =(v/c)B$ which takes into account that the voltage drop occurs in the
stripe only.

In the viscous regime of flux flow the vortex velocity is proportional to the Lorentz force, which is proportional to the current density $J$ in the film. The vortex velocity is defined by the stationary limit of the equation of motion of a single vortex
\begin{equation}
    \label{2}
    \eta(v) v = (1/c)J\phi_0,
\end{equation}
where $\eta(v)$ is the nonlinear viscosity coefficient \cite{Lar75etp,Lar86inb}
\begin{equation}
    \label{3}
    \eta(v)= \eta(0)\Bigl[1+\Bigl(\frac{v}{v^\ast}\Bigr)^2\Bigr]^{-1},
\end{equation}
where
\begin{equation}
    \label{4}
    \eta(0) = 0.45 \frac{\sigma_\mathrm{n} T_\mathrm{c}}{D}\sqrt{1-\frac{T}{T_\mathrm{c}}}
\end{equation}
and
\begin{equation}
\label{5}
    v^\ast = 1.02 (D/\tau_{\varepsilon})^{1/2}(1-T/T_\mathrm{c})^{1/4}.
\end{equation}
In Eqs. (\ref{4}) and (\ref{5}) $\sigma_\mathrm{n}$ is the conductivity of the film in the normal state, $\tau_{\varepsilon}$ is the quasiparticle energy relaxation time, and $D$ is the quasiparticle diffusion
coefficient.

As justified in previous works \cite{Lar75etp,Lar86inb}, we assume that the diffusion length $l_{\varepsilon}=(D\tau_{\varepsilon})^{1/2}$ is much larger than the size of the vortex core, $l_{\varepsilon}\gg\xi(T)$ . In addition, in what follows we will assume $l_{\varepsilon} \ll \delta$, that is the diffusion length is much smaller than the stripe width.

The quasiparticle temperature $T$ entering Eqs. (\ref{4}) and (\ref{5}) is inhomogeneous along the $x$-coordinate and it obeys the heat conduction equation
\begin{equation}
    \label{6}
    k\frac{d^2 T}{dx^2}=\frac{h(T_\mathrm{B})}{d}(T -T_\mathrm{B}) - \frac{IV}{wd}\delta(x).
\end{equation}
In this equation, $k$ is the heat conduction coefficient, $h(T_B)$ is the heat removal coefficient taken at the substrate temperature $T_\mathrm{B}$ and $I$ is the current flowing in the film. The heat conduction equation with a heat source proportional to the $\delta$-function is justified when $l_\mathrm{T}= \sqrt{kd/h} \gg \delta$, i.e. the typical length scale of the temperature variation is much larger than the width of the stripe where heat is released.

The solution of Eq. (\ref{6}) reads
\begin{equation}
    \label{7}
    T(x) = \frac{IVl_\mathrm{T}}{2kwd}e^{-|x|/l_\mathrm{T}}+T_\mathrm{B}.
\end{equation}
In the limiting case $\delta\ll l_\mathrm{T}$ one can treat the stripe temperature as equal to
\begin{equation}
    \label{8}
    T(0) = \frac{IV}{2w\sqrt{hkd}}+T_\mathrm{B}.
\end{equation}
This expression for $T(0)$ will be used in the following analysis.

\subsection{Local flux-flow instability for finite heat removal rate}
\label{s3}

The flux-flow instability is caused by the nonlinear dependence of the film conductivity on the electric field $\sigma (E)$. The expression for $\sigma (E)$ obtained by Larkin and Ovchinnikov \cite{Lar75etp} in the dirty limit near $T_\mathrm{c}$ reads
\begin{equation}
    \label{24}
\sigma
(E)=\sigma_\mathrm{n}\frac{H_\mathrm{{c2}}(T)}{B\sqrt{1-T/T_\mathrm{c}}}\frac{f(B/H_\mathrm{{c2}})}{1+(E/E^\ast)^2},
\end{equation}
Here, $\sigma_\mathrm{n}$ is the conductivity of the film in the normal state, $E$ is  the electric field, and $E^\ast = v^\ast B/c$ with $v^\ast$ being the vortex velocity at the instability point. The function $f(B / H_\mathrm{{c2}})$ appears due to the overlap of vortex cores and $f (B / H_\mathrm{{c2}}) \approx 4.04$ for magnetic fields of interest here, which are $B \lesssim 0.4H_\mathrm{{c2}}$ \cite{Lar86inb,Pru90pcs}. We refer to Ref. \cite{Lar86inb} where $f (B / H_\mathrm{{c2}})$ is tabulated for a wider range of magnetic field values.

It should be stressed that the quasiparticle temperature $T$ entering Eq. (\ref{24}) depends on the electric field and therefore on the vortex velocity. Introducing the parameters $T^\ast$ and $E^\ast$ for the temperature and the electric field corresponding to the instability point one can write a system of the heat balance equation derived from Eqs. (\ref{2})--(\ref{7})
\begin{equation}
    \label{25}
    T-T_\mathrm{B}=\frac{B\delta }{2\phi_0 \sqrt{hkd}}\eta(v)v^2
\end{equation}
combined with the extremum condition in the $I$-$V$ curve
\begin{equation}
    \label{26}
    \frac{d}{dE}[\sigma(E)E]_{E=E^\ast} =0.
\end{equation}
Introducing the dimensionless variables $e = E^\ast / E^{\ast} (T_\mathrm{B})$  and $t =
(T_\mathrm{c} -T^\ast) / (T_\mathrm{c} -T_\mathrm{B})$, and using the expression
\begin{equation}
    \label{27}
    H_\mathrm{{c2}}(T)=\frac{4\phi_0}{\pi^2 \hbar D}k_\mathrm{B} (T_\mathrm{c} -T)
\end{equation}
for the upper critical field, which is justified for superconductors with a short mean free  path
of quasiparticles, the system of Eqs. (\ref{25}) and (\ref{26}) can be rewritten as
\begin{equation}
    \label{28}
    1-t=2bte^2/(e^2+\sqrt{t}),
\end{equation}
\begin{equation}
    \label{29}
    1+\frac{e}{2t}\frac{dt}{de}-\frac{e^2}{\sqrt{t}}\Bigl(1-\frac{e}{t}\frac{dt}{de}\Bigr)=0,
\end{equation}
where $b = B / B_\mathrm{T}$ is the dimensionless magnetic field with the parameter
\begin{equation}
    \label{30}
    B_\mathrm{T} = 0.374 k_\mathrm{B}^{-1}c e_0 R_{\square}h\tau_\varepsilon(2l_\mathrm{T}/\delta).
\end{equation}
Here, $R_{\square} = (\sigma_n d)^{-1}$ is the film resistance per square. The value of $B_\mathrm{T}$ in Eq. (\ref{30}) is by the factor $(2l_\mathrm{T}/\delta)\gg1$ larger than $B_\mathrm{T}$ in the homogenous case
\cite{Bez92pcs}. It can be shown that for an arbitrary value of $(2l_\mathrm{T}/\delta)$, the parameter $B_\mathrm{T}$ reads
\begin{equation}
    \label{30a}
    B_\mathrm{T} = 0.374 k_\mathrm{B}^{-1}c e_0 R_{\square}h\tau_\varepsilon (1-e^{-\delta/2l_\mathrm{T}})^{-1}.
\end{equation}
\begin{figure*}[t!]
    \includegraphics[width=0.62\linewidth]{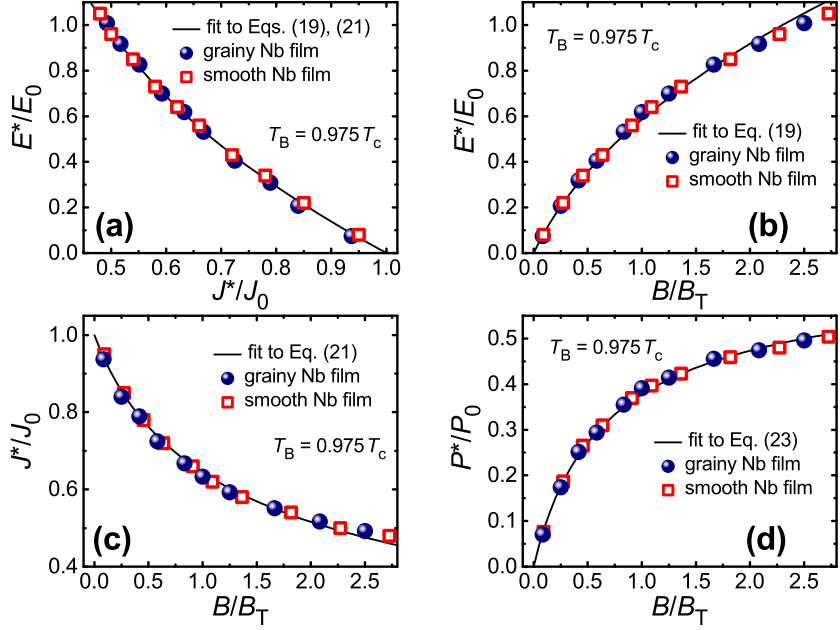}
\caption{(a) The complete set of instability points described by Eqs. (\ref{33}) and (\ref{35}).
Dependences of the electric field $E^\ast$ (b), current density $J^\ast$ (c) and specific power
$P^\ast$ (d) at the instability points on the normalized magnetic field. Symbols are experimental
data for two Nb films at $T_\mathrm{B}=0.975T_\mathrm{c}$. Spheres: Nb film with grainy
morphology. Squares: Nb film with smooth morphology. Solid lines: calculations by Eqs. (\ref{33}),
(\ref{35}) and (\ref{37}) with
    $J_0= 49.7$\,kA/cm$^2$,
    $E_0 = 0.12$\,V/cm,
    $P_0 = E_0 J_0$ and
    $B_\mathrm{T} =12$\,mT
    for the Nb film with grainy morphology and
    $J_0= 42.8$\,kA/cm$^2$,
    $E_0 = 0.2$\,V/cm,
    $P_0 = E_0 J_0$ and
    $B_\mathrm{T} =11$\,mT for the Nb film with smooth morphology.
    }\label{fig2}
\end{figure*}

At small $\delta/l_\mathrm{T}$ Eq. (\ref{30a}) reduces to Eq. (\ref{30}) while at large $\delta/l_\mathrm{T}$ the results of Ref. \cite{Bez92pcs} for homogeneous flux flow are reproduced. The system of Eqs. (\ref{28}) and (\ref{29}) reads as in the homogenous case
\cite{Bez92pcs}. Its solution is
\begin{equation}
    \label{31}
    t=[1+b+(b^2 +8b+4)^{1/2}]/3(1+2b),
\end{equation}
\begin{equation}
    \label{32}
    e^2=(1/2)\sqrt{t}(3t-1).
\end{equation}
From Eqs. (\ref{31}) and (\ref{32}) one obtains the following expression for the instability electric field $E^\ast$
\begin{equation}
    \label{33}
    \frac{E^\ast}{E_0}=\frac{(1-t)(3t+1)}{2\sqrt{2}t^{3/4}(3t-1)^{1/2}}
\end{equation}
where
\begin{equation}
    \label{34}
    E_0 = 1.02(B_\mathrm{T}/c)(D/\tau_\varepsilon)^{1/2}(1-T_\mathrm{B}/T_\mathrm{c})^{1/4}
\end{equation}
is independent of the magnetic field but depends on the width $\delta$ of the stripe with  moving vortices via Eq. (\ref{30a}).

The instability current density $J^\ast = \sigma(E^\ast) E^\ast$ reads
\begin{equation}
    \label{35}
    \frac{J^\ast}{J_0} = \frac{2\sqrt{2}t^{3/4}(3t-1)^{1/2}}{3t+1},
\end{equation}
where
\begin{equation}
    \label{36}
J_0 = 2.62 (\sigma_\mathrm{n}/e_0)(D\tau_\varepsilon)^{-1/2}k_\mathrm{B} T_\mathrm{c}
(1-T_\mathrm{B}/T_\mathrm{c})^{3/4}
\end{equation}
corresponds to $J^\ast$ at $B = 0$.

Expressions (\ref{33}) and (\ref{35}) describe the comprehensive set of all instability points $E^\ast (J^\ast)$ in the $I$-$V$ curves acquired at different values of the magnetic field at a given substrate temperature $T_\mathrm{B}$. The dependence $E^\ast = E^\ast (J^\ast)$ calculated from Eqs. (\ref{33}) and (\ref{35}) is plotted in Fig. \ref{fig2}(a) and should be compared with instability points deduced from experiment, which are also shown by the symbols in Fig. \ref{fig2} and will be discussed in what follows. The instability parameters $E^\ast$ and $J^\ast$ given by Eqs. (\ref{33}) and (\ref{35}) depend on the magnetic field through the parameter $t$ defined by Eq. (\ref{31}). The respective dependences are plotted in Figs. \ref{fig2}(b) and \ref{fig2}(c).

Finally, the specific power at the instability point is defined by
\begin{equation}
    \label{37}
    P^\ast/P_0 = 1- t(b),
\end{equation}
where $t(b)$ is determined by Eq. (\ref{31})  and $P_0 = E_0 J_0 = (h/d)(T_\mathrm{c}-T_\mathrm{B})(1-e^{-\delta/2l_\mathrm{T}})^{-1}$. The dependence of $P^\ast/P_0$ on $B/B_\mathrm{T}$ is plotted in Fig. \ref{fig2}(d). From the parameter $P_0$ one can deduce the effective heat removal coefficient $h_\mathrm{eff} = h(1-e^{-\delta/2l_\mathrm{T}})^{-1}$. Then, by substituting $h_\mathrm{eff}$ into Eq. (\ref{30a}) for $B_\mathrm{T}$ the inelastic quasiparticles scattering time $\tau_\varepsilon$ can be deduced.

\begin{figure*}[tbh!]
    \includegraphics[width=0.62\linewidth]{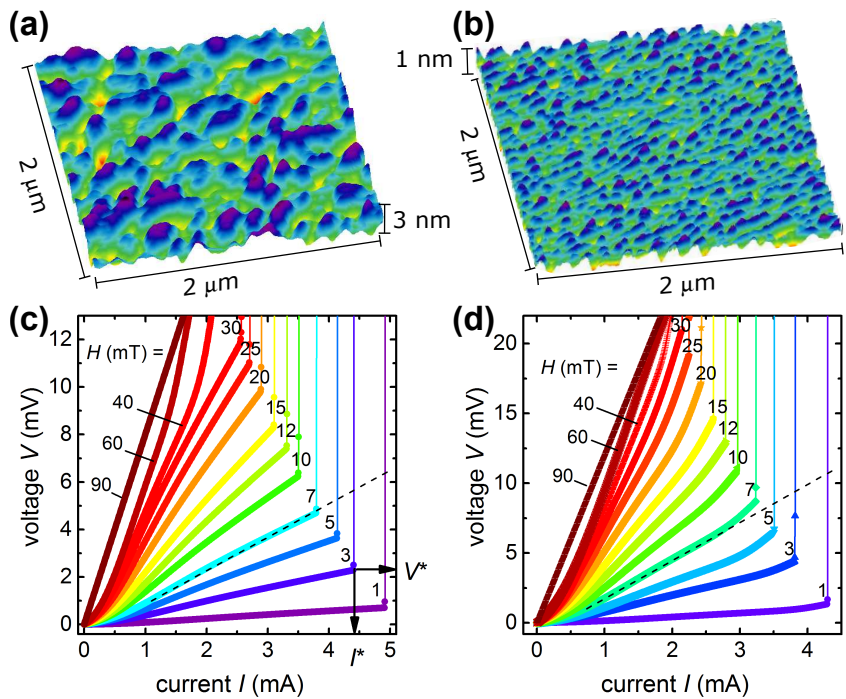}
    \caption{Atomic force microscopy images of the surface of the Nb film with grainy (a) and smooth (b) morphology. The measured $I$-$V$ curves for the two Nb films at $T_\mathrm{B} = 0.975T_\mathrm{c}$ are shown in (c) and (d), respectively. The dashed lines are guides to the eye emphasizing that the instability jumps occur in the \emph{linear} $I$-$V$ sections in (c) and \emph{nonlinear} $I$-$V$ sections in (d). The determination of the instability parameters $V^\ast$ and $I^\ast$ is indicated. No instability jumps are observed at $H \gtrsim 30$\,mT.
    }\label{fig3}
\end{figure*}

\section{Experiment}
\label{s4}
In order to examine the theoretical model, $I$-$V$ curves were measured on two Nb films grown with different substrate temperatures resulting in different pinning conditions for Abrikosov vortices. The films are 70\,nm-thick epitaxial (110) Nb films sputtered by dc magnetron sputtering on a-cut sapphire substrates. The films were sputtered in a setup with a base pressure in the $10^{-8}$\,mbar range. In the sputtering process the substrate temperature was $850^\circ$C, the Ar pressure was $4\times10^{-3}$\,mbar, and the growth rate was about $1$\,nm/s. One film was deposited on a sapphire substrate heated to $850^\circ$, while the substrate temperature was $600^\circ$ during the deposition of the second film. The substrate temperature and the deposition rate affect the microstructural properties and the pinning strength in the films \cite{Wil01tsf,Dob12tsf}. Accordingly, the first film exhibits a smooth morphology while a grainy morphology resulted for the second film, respectively. The rms surface roughness of the film with grainy morphology is about $3$\,nm, as deduced from inspection by non-contact atomic force
microscopy over a scan range of $2\,\mu$m$\times2\,\mu$m, see Fig. \ref{fig3}(a) and (b). The film deposited with a higher substrate temperature has a smoother surface with an rms roughness of about $1$\,nm. X-ray diffraction measurements revealed the (110) orientation of the films and the epitaxy of the films has been confirmed by reflection high-energy electron diffraction \cite{Dob12tsf}.

\vspace{5mm}
For electrical resistance measurements $0.15\times1$\,mm$^2$ bridges were fabricated by photolithography and Ar ion-beam etching of both films. The films show superconducting transition temperatures $T^\mathrm{grainy}_\mathrm{c}=8.86$\,K and $T^\mathrm{smooth}_\mathrm{c}=8.98$\,K, as deduced by a $50$\% resistance drop criterion. The upper critical field $H_\mathrm{c2}(0)$ of both films is about $1.1$\,T, as deduced from fitting the dependence $H_\mathrm{c2}(T)$ to the phenomenological law $H_\mathrm{c2}(T) = H_{c2}(0) [1-(T/T_\mathrm{c})^2]$. The values of the superconducting coherence length $\xi(0)$ deduced from the relation $\xi(0) = [\Phi_0/(2\pi H_\mathrm{c2}(0))]^{1/2}$ were found to be around $17$\,nm for both films.

The electrical resistance measurements were done with magnetic field oriented perpendicular to the film surface. To minimize self-heating effects, $I$-$V$ curves were measured in a pulsed current-driven upsweep mode with a rectangular pulse width of $1$\,ms and a pulse-off time of $1$\,s. The dissipated power $P^\ast = I^\ast V^\ast$ at different magnetic fields was derived from the currents $I^\ast$ and voltages $V^\ast$ at the instability points, refer to Fig. \ref{fig2}(d). In the case of Joule overheating leading to thermal runaway, $P$ is expected to be independent of the magnetic field. Since this was not the case in our experiments, one can rule out that the observed flux flow instability points relate to the Joule thermal runaway effect \cite{Xia99prb}.

The $I$-$V$ curves were measured at $T_\mathrm{B} = 0.975T_\mathrm{c}$ for a series of magnetic fields in the range $0$ to $90$\,mT. The measured $I$-$V$ curves for both films are shown in Fig. \ref{fig3}(c) and (d). For both samples the $I$-$V$ curves exhibit a dissipation-free regime at very small current densities and a nearly linear regime relating to viscous flux flow at current densities larger than the depinning current density. However, we emphasize the qualitatively different behavior of the $I$-$V$ curves of both samples at larger currents. Namely, in the film with grainy morphology the $I$-$V$ curves maintain their linearity up to the instability point. In contradistinction, an upward bending at the foot of the instability jumps is seen in all $I$-$V$ curves for the film with smooth morphology. From the last data point before the jump at $I^\ast$ the critical vortex
velocity $v^\ast$ was derived by the relation $v^\ast = c V^\ast/(B L)$, where $B$ is the applied magnetic field and $L=1\,$mm is the distance between the voltage contacts. The instability current density $J^\ast =I^\ast/wd$ is determined from the current $I^\ast$ relating to $V^\ast$. Here, $w = 150\,\mu$m and $d$ are the width and the thickness of the superconducting film. At larger magnetic fields $B\gtrsim30$\,mT, the $I$-$V$ curves become smooth and the instability jumps disappear altogether.

\section{Discussion}
\label{s5}

In the theoretical model, the most crude assumption, that has allowed us to solve the problem analytically, was that the stripe with moving vortices is located between film areas in which vortices are pinned. In particular, in the derivation of Eq. (\ref{30a}) it was assumed that the flux-flow instability appears in the region where the film temperature is maximal, that is in the middle of the stripe where the temperature is equal to $T(0)$. In actual fact, however, the situation is more complicated and there are channels in which vortices move faster and slower, as corroborated by numerical simulations based on the time-dependent Ginzburg-Landau equation \cite{Sil12njp,Ada15prb}. Previously, such stripe-like flux patterns were visualized by scanning Hall microscopy and they are sometimes termed as \emph{vortex rivers} \cite{Sil10prl,Sil10pcs}. Indeed, in consequence of variations of the local pinning forces in samples there are regions where vortices move almost freely, while a stronger pinning in other regions leads to slower motion of vortices or their local anchoring.

A broad distribution of vortex velocities caused by the presence of regions with different pinning strengths has an important consequence. Namely, on average, the essentially larger number of slowly moving vortices makes a larger contribution to the measured $I$-$V$ curve as compared to the contribution of the much smaller number of faster moving vortices. Therefore, the $I$-$V$ curve in this case maintains a linear shape up to the instability point. The local flux-flow instability occurs upon reaching the instability threshold current in areas with weaker pinning. These vortices provide a much smaller resistance contribution due to their small number. In consequence of the overheating of the local areas of faster moving vortices, a normal domain can be formed across the superconducting film. Whether this domain will vanish or grow depends on the relation between the instability current and the current of equilibrium of a non-isothermic N/S boundary. Namely, for $I_0 > I_\mathrm{eq}$ the normal domain grows and the entire film transits into the normal state. In this  case, the (almost) linear section of the $I$-$V$ curve, which are often observed experimentally \cite{Vol92fnt,Gri11snm} terminates at the instability point above which the film transits into the normal state. For completeness, we note that in the absence of strong pinning sites or in the case of rather small variations of pinning forces along the trajectories of moving vortices, the $I$-$V$ curve is essentially nonlinear. In this case, the last point before the jump in the $I$-$V$ curve corresponds to the instability occurring in a large region of the film. Such $I$-$V$ curves were also observed experimentally \cite{Mus80etp,Kle85ltp,Vol92fnt,Xia99prb,Per05prb,Att12pcm,Sil12njp,Gri15prb}.
\begin{figure}[t!]
    \includegraphics[width=0.68\linewidth]{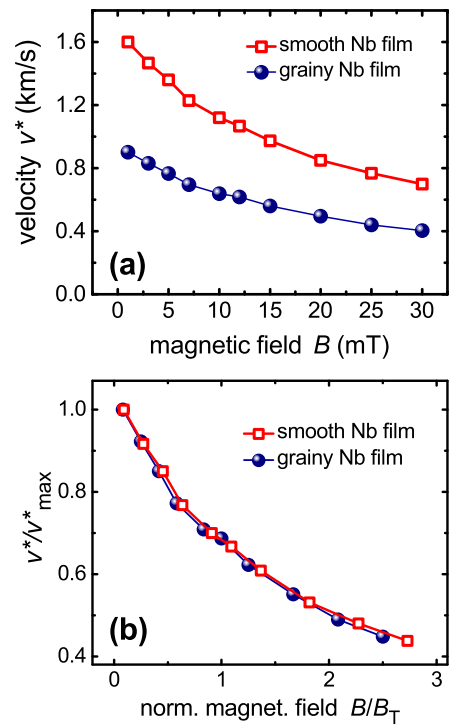}
    \caption{(a) The deduced instability velocities $v^\ast(B)$ as a function of  the magnetic field for the grainy and smooth Nb films at $T = 0.975 T_\mathrm{c}$. (b) The same data as in (a) but in the normalized $v^\ast/v^\ast_\mathrm{max}$ versus $B/B_\mathrm{T}$ representation.
    }\label{fig4}
\end{figure}

In general, in the case of local instability occurring in a region of width $\delta$, it would seem that one should use the relation $v^\ast =c V^\ast/(B \delta)$ rather than $v^\ast = c V^\ast/(B L)$. However, within the framework of this approach, the $I$-$V$ curve will be nonlinear, which contradicts with the experimental data for the Nb film with grainy morphology. At the same time, if we deduce the vortex velocity by the standard expression $v^\ast = c V^\ast/(B L)$ then we get $v^\ast$ values which are a factor of about two smaller than for the smooth film, as depicted in Fig. \ref{fig4}(a). Remarkably, if we plot the magnetic field dependence in the normalized
$v^\ast/v^\ast_\mathrm{max}$ versus $B/B_\mathrm{T}$ representation, see Fig. \ref{fig4}(b), then the magnetic field dependences of $v^\ast$ for both samples nicely coincide. Here, $v^\ast_\mathrm{max}$ is the value of $v^\ast$ at $1$\,mT.

As a check of the evolution of the normal domain, we make estimates for $J_\mathrm{eq}$ for  both samples by the relation
\begin{equation}
    \label{38}
    J_{eq} = \biggl[\frac{2 h}{R_\square d^2} (T_\mathrm{c} - T_\mathrm{B})\biggr]^{1/2}.
\end{equation}
Namely, for the film with smooth morphology at zero magnetic field we deduce $J_\mathrm{eq}\approx36$\,kA/cm$^2$, i.e. $J_\mathrm{eq}< J_0 = 42.8$\,kA/cm$^2$. For the film with grainy morphology we obtain $J_\mathrm{eq}\approx 38$\,kA/cm$^2$ so that the same inequality $J_\mathrm{eq} < J_0 = 49.7$\,kA/cm$^2$ holds for this sample as well. These inequalities mean that in both Nb films the normal domains grow across the entire superconducting film, which is an important check for the case of local instability.

We emphasize that the theory of local flux-flow instability developed in this paper should be applied to $I$-$V$ curves exhibiting a linear section (linear flux-flow regime) maintained up to the instability onset. A comparison of the theory with experimental results is shown in Figs. \ref{fig2} and \ref{fig3} where a good agreement between the experimental data and the calculations is revealed. From the experimental data we deduce a heat removal coefficient of $h_\mathrm{eff} = 0.27$\,WK$^{-1}$cm$^{-2}$ and an inelastic energy relaxation time of quasiparticles of $\tau_\varepsilon = 0.52 $\,ns for the film with smooth morphology. For the Nb film with grainy morphology we deduce $h_\mathrm{eff} = 0.18$\,WK$^{-1}$cm$^{-2}$ and $\tau_\varepsilon = 1.28$\,ns. We note that in Ref. \cite{Ger90etp} the value and the temperature dependence of the inelastic scattering time for electrons in Nb were measured directly. It turned out that both these characteristics depend strongly on the electron mean free path and the film thickness, we refer to Fig. 5 in Ref. \cite{Ger90etp}. At the same time, the deduced values $\tau_\varepsilon \simeq 1$\,ns in our experiments coincide in the order of magnitude with the
values deduced in Refs. \cite{Ger90etp,Per05prb,Leo11prb}. We also note that the values of the quasiparticle relaxation time $\tau_{\varepsilon}$ obtained in the framework of the Larkin-Ovchinnikov theory are different from those estimated from photoresponse experiments, because of different excitation energies \cite{Kap76prb}. Yet, in recent work \cite{Cap17apl} it was pointed out that the scaling between the $\tau_{\varepsilon}$
values extracted within the vortex instability approach and optical experiments are the same for NbN and Nb thin films \cite{Par05sst,Cir11prb}. For this reason, flux-flow instability studies are highly relevant for the characterization of candidate materials for fast single-photon detection \cite{Cap17apl} and therefore understanding of mechanisms of the flux-flow instability is pivotal for the optimization of the superconducting devices' performance.

In general, while the quantitative agreement of our theory with experiment corroborates the LO mechanism of the flux-flow instability in the investigated system, the considered instability mechanism in a current-carrying state of superconducting films is not the only one. For instance, in recent work \cite{Qia18nsr} an instability mechanism associated with the generation of free vortices in consequence of a Berezinskii-Kosterlitz-Thouless (BKT) transition was considered. The BKT instability was argued to occur at $T_\mathrm{B}\leq T_\mathrm{BKT}$ and it gets very quickly suppressed by the magnetic field \cite{Qia18nsr}. This is why this mechanism is very unlikely in our experiments at $T_\mathrm{B} \approx T_\mathrm{c}$ and $0.01H_\mathrm{c2}(T_\mathrm{B})\lesssim H \lesssim 0.5 H_\mathrm{c2}(T_\mathrm{B})$.

Finally, in contradistinction with the standard LO instability scenario in the presence of magnetic fields near $T_\mathrm{c}$, there is a further hot-electron mechanism \cite{Kun02prl,Kni06prb} dominating at low temperatures $T_\mathrm{B}\ll T_\mathrm{c}$. In this, the main effect of the dissipation is to raise the electronic temperature, create additional quasiparticles, and diminish the order parameter \cite{Kun02prl,Kni06prb}. In contradistinction with the LO mechanism, the vortex expands rather than shrinks, and the viscous drag is reduced because of a softening of gradients of the vortex profile rather than a removal of quasiparticles. We note that the effect of pinning on the hot-electron flux-flow instability was analyzed theoretically \cite{Shk17pcs} and allowed for fitting a non-monotonic magnetic-field dependence of the instability velocity in Nb thin films with different pinning strengths \cite{Dob17sst}.

\section{Conclusion}
\label{s6}

In summary, in this work an analysis of the local flux-flow instability in superconducting  thin films has been presented with account for a finite rate of heat removal into the substrate. The main distinctive feature of the problem considered here is the assumption that the flux-flow instability occurs not in the entire film but in a small region represented by a stripe of length $\delta$ across of the film width. The local character of the instability leads to the replacement of the heat removal coefficient $h$ by an effective heat removal coefficient $h_\mathrm{eff}$ which depends on the relation of $\delta$ to the characteristic length scale $l_\mathrm{T}$ of the temperature variation. This replacement essentially extends the applicability of the flux-flow instability model considered so far only in the homogeneous case, and it allows one to deduce the inelastic quasiparticle relaxation time $\tau_\varepsilon$. At the same time, the dependences of $E^\ast/E_0$ on $J^\ast/J_0$ and of both  $E^\ast/E_0$ and $J^\ast/J_0$ on $B/B_T$ obtained within the framework of the theory developed in Ref.~\cite{Bez92pcs} remain
universal and allow one to describe experimental data by varying $E_0$ and $B_T$ as two fitting parameters.

\begin{acknowledgments}
This work was partially supported by the MES of Ukraine through Project No. 0118U002037.
OD acknowledges the German Research Foundation (DFG) for support through Grant No 374052683 (DO1511/3-1).
This work was also supported by the European Cooperation in Science and Technology via COST Action CA16218 (NANOCOHYBRI).
\end{acknowledgments}


%

\end{document}